\documentclass{worldsci}       

\begin{document}

\title{Layer-by-layer pattern propagtion \\ and pulsed laser deposition}

\author{F.\ Westerhoff, L.\ Brendel, D.E.\ Wolf}

\address{Gerhard-Mercator-Universit\"at Duisburg, 47048 Duisburg}


\maketitle

\abstracts{ In this article kinetic Monte Carlo simulations for molecular beam
  epitaxy (MBE) and pulsed laser depositon (PLD) are compared. It will be shown
  that an optimal pattern conservation during MBE is achieved for a specific
  ratio of diffusion to deposition rate. Further on pulsed laser deposition is
  presented as an alternative way to control layer by layer growth. First
  results concerning the island density in the submonolayer regime are shown.}

\section{Introduction}
Multi-layered structures with sharp interfaces are of
high interest, due to the new physical properties they exhibit resulting from
the reduced dimensionality. 
In order to obtain such sharp interfaces the deposited material has to grow
in layers instead of forming 3D islands.
Two different methods of growing thin films will be compared here.

Molecular beam epitaxy (MBE) allows to produce mono-crystalline layer
structures with high sharpness at the interfaces. Growth takes place in an
ultra high vacuum chamber in which the material is thermally evaporated and thus
is deposited on the substrate with a continuous flux $F$.

The fabrication of thin films with pulsed laser deposition (PLD) is a quite new
technique which can improve layer by layer
growth.\cite{kirschner1996} In addition PLD provides the possibility to deposit
any desired material with the stoichiometry of the ablated material being conserved.
In contrast to thermal evaporation in MBE many particles are ablated from the target 
simultaneously, resulting in a high adatom density
directly after the pulse. This causes an increased islands
density in the PLD as will be described. 

\section{Molecular Beam Epitaxy}
For an initially flat surface layer by layer growth usually improves, the
higher the temperature of the substrate is.\cite{kallaJMP1997} But 
a pattern (e.g. a structure produced by lithographic techniques) will
in general not be conserved better during layer by layer growth, if
the substrate temperature is raised. The reason is, that for high
temperatures the adatoms do not
nucleate on top of the islands any more because the diffusion
length becomes too large. Instead they are incorporated at the edges of the islands and therefore
flatten the surface which means that information about the pattern
gets lost. 
As the diffusion length depends on the ratio of the diffusion rate to the deposition
rate,\cite{wolf1997} this value has to be optimized if a given
pattern should be conserved during growth. One can also consider the
opposite optimization problem: How to choose the growth parameters in
order to heal any unwanted structures as efficiently as possible? As
the two optimization problems are obviously two sides of the same
medal, we consider only the first one in the following.

\subsection{Model}
In the simulation we use a simple cubic crystal with periodic boundary
conditions and assume that growth takes place without overhangs and defects.
The size of the substrate is $(L \cdot a)^2$, $L^2$ denoting the number of lattice
sites and $a$ the lattice constant. 
The atoms are deposited with a flux $F$ and diffuse with the diffusion constant
$D$ until they are irreversibly bound to another adatom or an island edge.

By {\em pattern propagation} we mean the deterministic height increase
by one lattice constant during  the deposition of one monolayer.
The probability for pattern propagation\cite{kallaPRL1997} is then the
fraction of sites which were conserved in every monolayer until time $t$:

\begin{equation}
p(t) =\left \langle \prod_{s=1}^{t} \delta_{h(\vec x,s),h(\vec x,s-1)+1} \right \rangle
\label{EFt}
\end{equation}

\subsection{Results}
\begin{figure}[t]
\begin{center}
\epsfxsize=28pc 
\epsfbox{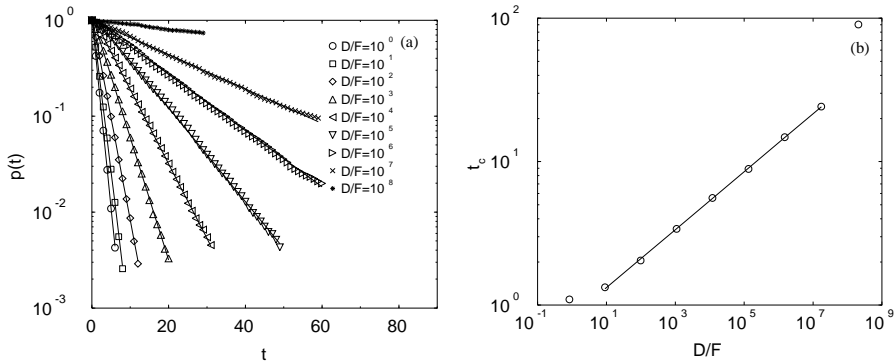} 
\caption{(a) Propagation for an initially flat surface. 
(b) Life time $t_c(D/F)$.\label{Fglatt65}}
\end{center}
\end{figure}

Figure \ref{Fglatt65} (a) shows $p(t)$ for an initially flat surface.
The exponential decay \mbox{$p(t) = \exp(-t/t_c)$} is what one would
expect, if the propagation of the surface sites from one layer to
the next would be temporally uncorrelated with constant transition
probability $p(1)$. Surprisingly, the exponential decay of $p(t)$ is
found although there is clear evidence that transition probabilities
are temporally correlated.\cite{dipl}
The {\em life time} $t_c$ (Fig.~\ref{Fglatt65}~b) of the initial
surface configuration is determined from the slope of $\ln(p(t))$.
The data give a power law for $t_c$ extending over 6 decades:
\begin{equation} 
t_c \propto (D/F)^{0.20 \pm 0.01}
\label{Etcglatt}
\end{equation}
In the limit of $D/F\rightarrow 0$ (random deposition) we get $t_c =1
$.\cite{kalladiss}  That the last data point in Fig.\ref{Fglatt65} (b)
deviates from the power law is a finite size effect.

In a  study of the one dimensional case\cite{kallaPRL1997,kalladiss} a logarithmic
dependence \mbox{($t_c \sim \log(D/F)$)} was found. Most recent
calculations seem to show a cross over from power law to logarithmic dependence
for larger $D/F$ in two dimensions, as well.\cite{WBLW}

\subsection{Structured surfaces}
\begin{figure}[t]
\begin{center}
\epsfxsize=28pc
\epsfbox{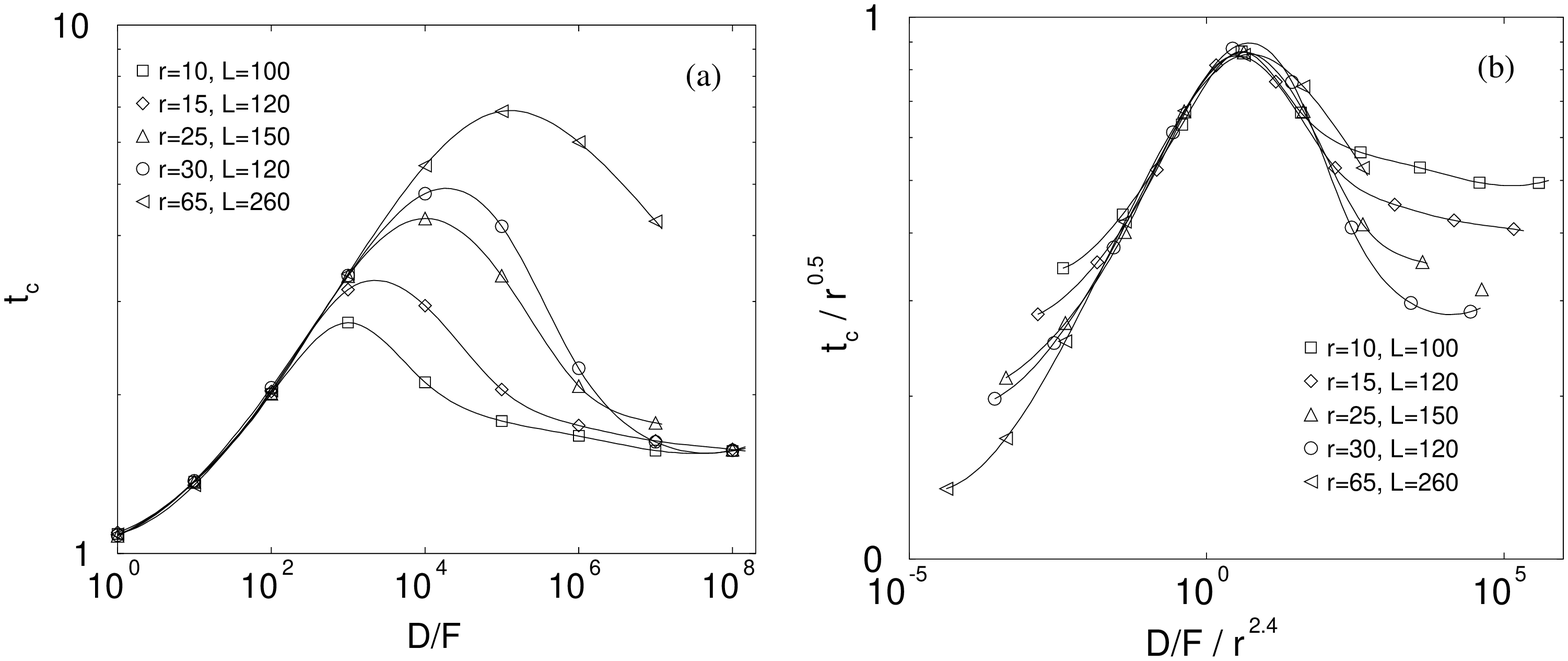}
\caption{(a) Life time $t_c$ for patterned surfaces. (b) Scaled
  data.\label{Ftc_spline}} 
\end{center}
\end{figure}

As long as the pattern is large compared to the diffusion length for an
initially flat surface, the propagation probability should not change since
the adatoms will not notice the pattern during their life time. Therefore we
also expect an exponential decay for small ratios of $D/F$. 
Here we  consider larger values for $D/F$, though, where we observe a different
behavior. We find a maximum in $t_c(D/F)$ which is shifted to higher $D/F$ with
growing pattern size $r$ (Fig.~\ref{Ftc_spline}~a). This can be
explained by  the existence of two different mechanisms
responsible for the pattern's decay. 

For small $D/F$ the life time $t_c$ is independent of the island size.
 This is the regime where the pattern vanishes through
kinetic roughening.  To the right of the maximum the diffusion length $l_D$ is
large compared to the pattern size, hence there are almost no nucleations on top
of the ``artificial'' islands. Instead, they take on the form of natural growing
and coalescing islands. In this way the pattern is also lost.

We find (see Fig.\ref{Ftc_spline} b) that the maximum life time
$t_{c,{\rm max}}$ for a given pattern length scale $r$ is reached for
\begin{equation}
(D/F)_{\rm max} \propto r^{2.4 \pm 0.1} \quad {\rm and \, scales\, like}\quad
t_{c,{\rm max}} \propto r^{0.50 \pm 0.05}.
\label{Emax}
\end{equation}
This is in agreement with (\ref{Etcglatt}): $t_{c,{\rm max}} \propto
(D/F)_{\rm max}^{0.5/2.4} = (D/F)_{\rm max}^{0.21}$. As (\ref{Etcglatt})
crosses over to a much weaker dependence for very large $D/F$, we
expect also that (\ref{Emax}) will be modified for larger $r$.

\section{Pulsed laser deposition}
In pulsed laser deposition the atoms are deposited with a pulse intensity $I$
(number of atoms per unit area) and frequency $\omega$ forming an effective
(average) flux 
\begin{equation}
F = I \cdot \omega\,.
\label{EF_eff}
\end{equation}

  If we reduce the intensity $I$ to the
limiting case of one atom per pulse and at the same time increase the frequency
so that the effective flux stays constant, we will reach conditions equivalent
to thermal deposition (MBE).\footnote{Nevertheless in the simulation there is still
  a small difference since in PLD atoms are deposited at fixed time intervalls,
  whereas in MBE fluctuations may occur.} One should therefore expect that both
methods show similar characteristics in the submonolayer regime.

If the intensity is increased, one should expect a higher island density,
because the adatom density and therefore the probability for nucleation
is also larger.  Hence we predict a {\em critical pulse intensity} $I_c$ which
separates both regimes.

In the simulation the kinetic energy of the arriving atoms is
neglected. This should be a valid approximation for
experiments, in which the laser intensity is just above the ablation
threshold\cite{kirschner1996} or the deposition takes place, e.g. in an oxygen
atmosphere which slows down the arriving particles.

\subsection{Critical pulse intensity}
During thermal deposition the time evolution of the adatom density
$\rho$ is given by a gain term, the flux $F$ and a loss term
$\rho/\tau$, where $\tau$ is the life time of adatoms on the surface:
\begin{equation}
\dot\rho = F-\frac{\rho}{\tau}\, .
\label{Erhopunkt}
\end{equation}
In the quasistationary regime (after island nucleation and before
island coalescence) the life time $\tau$ is determined by the typical
distance $l_D$ an adatom diffuses before reaching an island edge\cite{villain}:
$\tau={l_D}^2/D$. Then $\dot \rho$ can be neglected and one gets:
\begin{equation}
\rho \approx F\cdot \tau= F \cdot \frac{{l_D}^2}{D}.
\end{equation}
With $l_D\propto\left( \frac{D}{F} \right)^\gamma$ and $\gamma=1/6$ for two dimensions
and irrevsible aggregation\cite{wolf1997} this leads to
\begin{equation}
\rho\propto{l_D}^{-4}.
\label{Erho_von_lD}
\end{equation}

Equation (\ref{Erho_von_lD}) describes the adatom density for thermal
deposition. In PLD with a pulse intensity $I$ much smaller than this
density, the adatom density will oscillate around this value with the
frequency $\omega$. Then one expects the same scaling of the island
density as for thermal deposition. By contrast a much larger pulse intensity
would prevent the adatom density in PLD to become on average similar to the
one in MBE. Hence the critical pulse intensity is expected to be of the order 
\begin{equation}
I_c \propto {l_D}^{-4} \propto \left( \frac{D}{F} \right)^{-2/3}.
\label{EIc_von_DF}
\end{equation}

In the case of a large pulse intensity, the island density increases to such a degree
that the life time of the adatoms is shorter than the time between two
pulses. Hence the adatom density and therefore the island density becomes
independent of the pulse frequency and only depends on $I$.

\subsection{Results}
\begin{figure}[t]
\begin{center}
\epsfxsize=28pc
\epsfbox{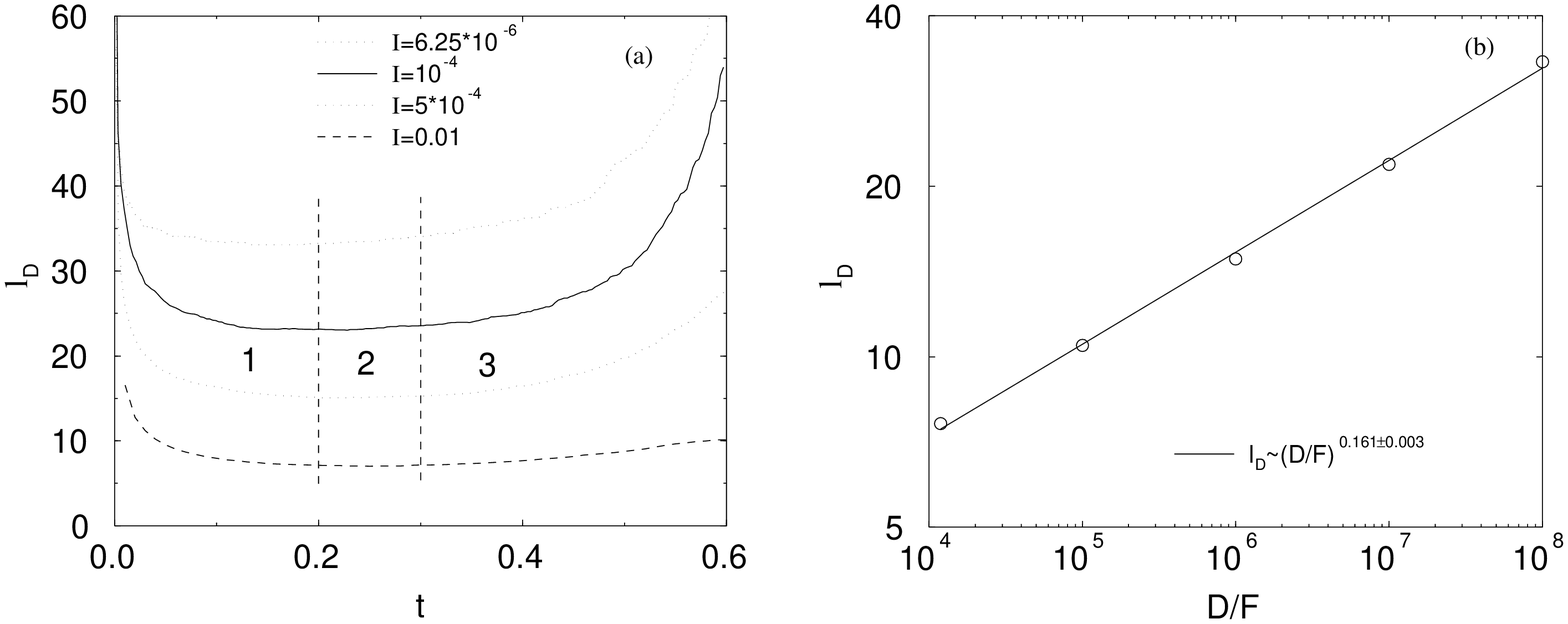}
\caption{(a) Island distance $l_D$ as a function of time $t$ (in units
  of monolayers). (b) $l_D(D/F)$ for the limiting case of one atom per pulse.
  \label{FPLD_lD_von_DF}}
\end{center}
\end{figure}

Fig.~\ref{FPLD_lD_von_DF}~(a) depicts the island density as a function of the
deposited monolayers. We expect three different regimes\cite{Thang1993}: The
regime for the  nucleation of islands, the saturation of the island density and
finally the coalescence of islands with an increasing island distance.
An increase of the pulse intensity leads to an overall decrease of the
island distance.

\subsection{$l_D$ for the limit of small pulse intensities}
According to a classical result, confirmed by experiments\cite{guenther1994} and
simulations\cite{Thang1993,stoyanov1991} the island distance scales like: 
\begin{equation}
l_D \propto \left(\frac{D}{F}\right)^{1/6}
\label{EJensen1}
\end{equation}
This behavior is confirmed in figure~\ref{FPLD_lD_von_DF}~(b) for the limit,
that only one atom per pulse ($I\cdot L^2 \rightarrow 1$) is deposited. 

\subsection{Scaling behavior of the island distance}
\begin{figure}[t]
\begin{center}
\epsfxsize=28pc
\epsfbox{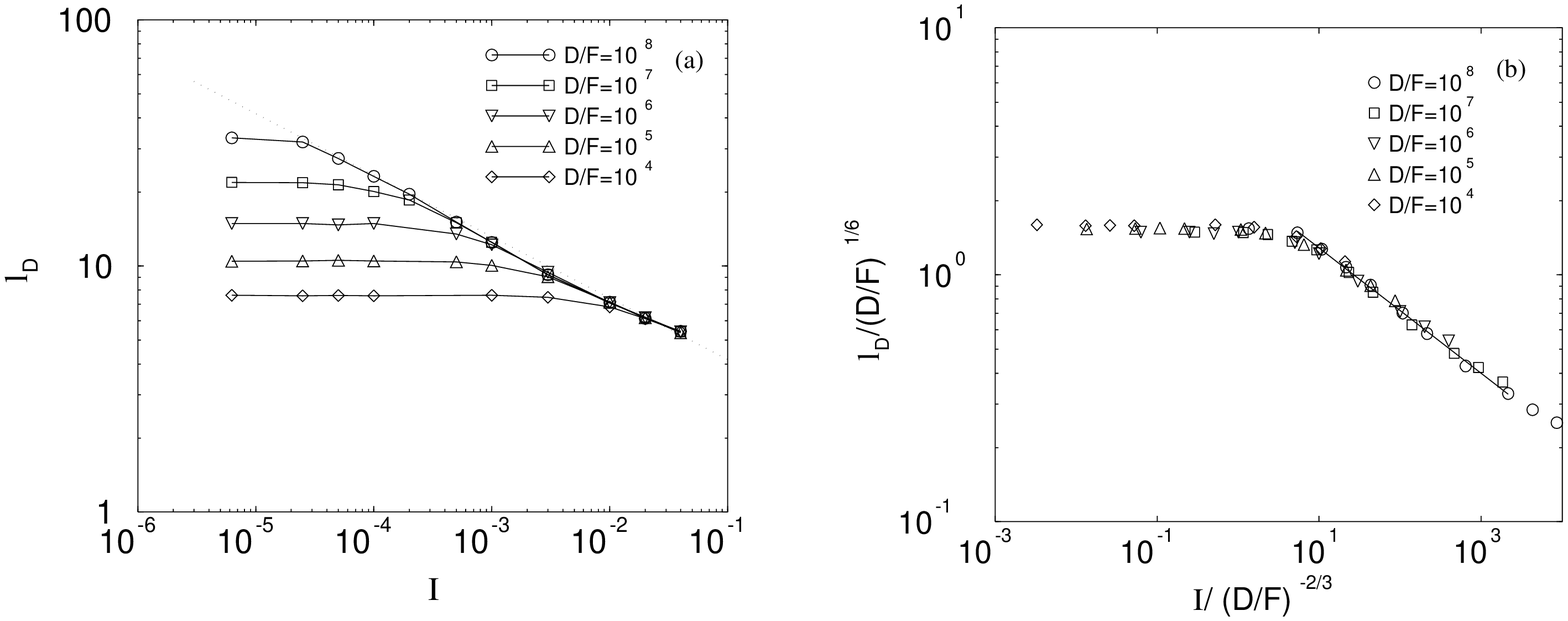}
\caption{(a) Island distance as a function of the pulse intensity for different
  ratios of $D/F$. (b) Scaled data.\label{FPLD_lD_von_I}} 
\end{center}
\end{figure}

Figure~\ref{FPLD_lD_von_I}~(a) depicts the island distance $l_D$ as a function
of the pulse intensity~$I$. For small values of $I$ the diffusion length
approaches a constant value, which is the island distance for the MBE case.
For larger pulse intensities $l_D$ is described by a power law:
\begin{equation}
l_D \propto I^{-\nu} \, .
\label{EJensen2}
\end{equation}
It turns out that $l_D$ can be written in a scaling form:
\begin{equation}
l_D\propto \left( \frac{D}{F}\right)^\gamma \cdot f\left( \frac{I}{I_c} \right)
\label{Escaling}
\end{equation}
with
\begin{equation}
f(y) \left\{ 
  \begin{array}{ll}
    = \mbox{const.} & \mbox{for } y\ll 1 \mbox{ and}\\
    \sim y^{-\nu}& \mbox{for } y \gg 1\, .\\
  \end{array}
\right.
\label{Ef_von_IIc}
\end{equation}
While the asymptotic power laws (\ref{EJensen1},\ref{EJensen2}) were
also obtained by Jensen and Niemeyer\cite{jensen}, the scaling law
(\ref{Escaling}) is new. In the previous work the pulses had a finite
duration in contrast to the case presented here. This additional
characteristic time spoiled the scaling.

\noindent
Since for large $I$, the $\omega$-dependence (via F (\ref{EF_eff})) has to vanish:
\begin{equation}
I_c^\nu \cdot \left( \frac{D}{F} \right)^{\gamma} = \mbox{const.} 
\end{equation}
and therefore:
\begin{equation}
I_c \propto\left(\frac{D}{F}\right)^{- \gamma/\nu} 
\end{equation}

\noindent
For large pulse intensities we find in Fig.~\ref{FPLD_lD_von_I}~(a) 
the slope $ \nu = 0.248$, so that, together with $\gamma=1/6$, we get:
\[
\gamma / \nu \approx 2/3,
\]
in accordance with the theoretical prediction (\ref{EIc_von_DF}).

\section*{Acknowledgments}
We thank the Neumann Institute of Computing (NIC) in
J\"ulich for providing computer time on the CRAY T3E for this project.

\end{document}